\documentclass[a4paper,fleqn,usenatbib]{mnras}   

\usepackage[T1]{fontenc}
\usepackage{ae,aecompl}
\usepackage{graphicx}	
\usepackage{amsmath}	
\usepackage{amssymb}	

\newcommand{\nvalone}{\textrm{N}\textsc{v}}
\newcommand{\civalone}{\textrm{C}\textsc{iv}}

\newcommand{\siiiialone}{[\textrm{Si}\textsc{iii}]}

\newcommand{\hi}{\textrm{H}\textsc{i}}
\newcommand{\honly}{\textrm{H}}
\newcommand{\oiii}{[\textrm{O}\textsc{iii}]}
\newcommand{\oii}{[\textrm{O}\textsc{ii}]}
\newcommand{\heiialone}{[\textrm{He}\textsc{ii}]}
\newcommand{\oiidoub}{[\textrm{O}\textsc{ii}]\ensuremath{\lambda3727,3729}}
\newcommand{\nev}{[\textrm{Ne}\textsc{v}]\ensuremath{\lambda3426}}
\newcommand{\auroral}{[\textrm{O}\textsc{iii}]\ensuremath{\lambda4364}}

\newcommand{\oiiiv}{[\textrm{O}\textsc{iii}]\ensuremath{\lambda5007}}

\newcommand{\oiiialone}{[\textrm{O}\textsc{iii}]}

\newcommand{\oiiidoub}{[\textrm{O}~\textsc{iii}]\ensuremath{\lambda\lambda4959,5007}}
 
\newcommand{\ha}{\ifmmode {\rm H}\alpha \else H$\alpha$\fi}
\newcommand{\hb}{\ifmmode {\rm H}\beta \else H$\beta$\fi}
\newcommand{\hg}{\ifmmode {\rm H}\gamma \else H$\gamma$\fi}
\newcommand{\lya}{\ifmmode {\rm Ly}\alpha \else Ly$\alpha$\fi}
\newcommand{\pg}{\ifmmode {\rm P}\gamma \else Pa$\gamma$\fi}
\newcommand{\lyb}{\ifmmode {\rm Ly}\beta \else Ly$\beta$\fi}
\newcommand{\lyg}{\ifmmode {\rm Ly}\gamma \else Ly$\gamma$\fi}
\newcommand{\ciiialone}{\textrm{C}\textsc{iii}]} 

\newcommand{\ciiidoub}{\textrm{C}\textsc{iii}]\ensuremath{\lambda\lambda1907,1909}}
\newcommand{\siii}{[\textrm{Si}\textsc{ii}]\ensuremath{\lambda1260}}

\newcommand{\cii}{[\textrm{C}\textsc{ii}]\ensuremath{\lambda1334}}

\newcommand{\nv}{\textrm{N}\textsc{v}\ensuremath{\lambda1240}}

\newcommand{\civ}{\textrm{C}\textsc{iv}\ensuremath{\lambda1548,1550}}

\newcommand{\heii}{\textrm{He}\textsc{ii}\ensuremath{\lambda1640}}
\newcommand{\oiiiuv}{\textrm{O}\textsc{iii}]\ensuremath{\lambda1661,1666}}

\newcommand{\flyc}{\ifmmode  \mathrm{f}_\mathrm{esc}\mathrm{(LyC)} \else $\mathrm{f}_\mathrm{esc}\mathrm{(LyC)}$\fi}

\def\kms{km s$^{-1}$}

\def\ergs{\ifmmode \mathrm{erg\hspace{1mm}s}^{-1} \else erg s$^{-1}$\fi}
\def\ergscm{erg s$^{-1}$ cm$^{-2}$}
\def\micron{\ifmmode \mu\mathrm{m} \else $\mu$m\fi}
\def\msun{\ifmmode \mathrm{M}_{\odot} \else M$_{\odot}$\fi}
\def\msunyr{\ifmmode \mathrm{M}_{\odot} \hspace{1mm}{\rm yr}^{-1} \else $\mathrm{M}_{\odot}$ yr$^{-1}$\fi}
\def\zsun{\ifmmode Z_{\odot} \else Z$_{\odot}$\fi}
\def\lsun{\ifmmode L_{\odot} \else L$_{\odot}$\fi}
\def\mstar{\ifmmode \mathrm{M}_{\star} \else M$_{\star}$\fi}

\title[Ionised channels at high redshift]
{Ionising the Intergalactic Medium by Star Clusters: The first empirical evidence}

\author [E.~Vanzella et al.]{
\parbox[t]{\textwidth}{E.~Vanzella$^1$\thanks{E-mail: eros.vanzella@inaf.it}\thanks{Based on observations collected at the European Southern Observatory under ESO programmes P0102.A-0391(A) and DDT-297.A-5012(A).},
G.~B.~Caminha$^2$,  F.~Calura$^1$, G. Cupani$^{3}$, M.~Meneghetti$^1$, M.~Castellano$^4$, P.~Rosati$^{5,1}$, A.~Mercurio$^6$, E.~Sani$^7$, C.~Grillo$^{8}$, R.~Gilli$^1$, M.~Mignoli$^1$, A.~Comastri$^1$, M.~Nonino$^3$, S.~Cristiani$^{3}$, M.~Giavalisco$^{9}$ and K.~Caputi$^{2}$
}
\vspace*{8pt}\\
$^1$INAF -- OAS, Osservatorio di Astrofisica e Scienza dello Spazio di Bologna, via Gobetti 93/3, I-40129 Bologna, Italy\\
$^2$Kapteyn Astronomical Institute, University of Groningen, Postbus 800, 9700 AV Groningen, The Netherlands\\
$^3$INAF -- Osservatorio Astronomico di Trieste, via G. B. Tiepolo 11, I-34143, Trieste, Italy\\
$^4$INAF -- Osservatorio Astronomico di Roma, Via Frascati 33, I-00078 Monte Porzio Catone (RM), Italy\\
$^5$Dipartimento di Fisica e Scienze della Terra, Universit\`a degli Studi di Ferrara, via Saragat 1, I-44122 Ferrara, Italy\\
$^6$INAF -- Osservatorio Astronomico di Capodimonte, Via Moiariello 16, I-80131 Napoli, Italy\\
$^7$European Southern Observatory, Alonso de Cordova 3107, Casilla 19, Santiago 19001, Chile \\
$^8$Dipartimento di Fisica, Universit\`a  degli Studi di Milano, via Celoria 16, I-20133 Milano, Italy\\
$^{9}$Astronomy Department, University of Massachusetts, Amherst, MA 01003, USA\\
}

\begin{document}
\date{}
\maketitle

\begin{abstract}
We present a VLT/X-Shooter spectroscopy of the Lyman continuum (LyC) emitting galaxy {\em Ion2} at z=3.2121 
and compare it to that of the recently discovered strongly lensed LyC$-$emitter at z=2.37, known as the {\em Sunburst}
arc. 
Three main results emerge from the X-Shooter spectrum: (a) the \lya\ has three distinct peaks with the central one 
at the systemic redshift, indicating a ionised tunnel through which both \lya\ and LyC radiation escape;
(b) the large O32 oxygen index (\oiiidoub\ / \oiidoub) of
$9.18_{-1.32}^{+1.82}$ is compatible to those measured in local (z $\sim 0.4$) LyC leakers; 
(c) there are narrow nebular high-ionisation metal lines with $\sigma_v < 20$ \kms, which confirms the presence of 
young hot, massive stars. The \heii\ appears broad, 
consistent with a young stellar component including Wolf$-$Rayet stars. 
Similarly, the {\em Sunburst} LyC$-$emitter shows a triple$-$peaked \lya\ profile and 
from VLT/MUSE spectroscopy the presence of spectral features arising from young hot and massive stars. 
The strong lensing magnification, ($\mu > 20$), suggests that this exceptional object is a gravitationally$-$bound star cluster observed at a cosmological distance,
with a stellar mass M $\lesssim 10^7$ \msun\ and
an effective radius smaller than $20$ pc. Intriguingly, sources 
like Sunburst but without lensing magnification might appear as {\em Ion2}$-$like galaxies, in which 
unresolved massive star clusters dominate the ultraviolet emission. This work supports 
the idea that dense young star clusters can contribute to the ionisation of the IGM
through holes created by stellar feedback. 
\end{abstract}


\section{Introduction}
Recently, extensive surveys attempting to identify and study galaxies emitting Lyman continuum (LyC)
radiation across a large range of cosmic time have yielded several low-redshift cases whose properties are believed to 
be representative of the galaxies at redshift $z>7$ that contributed the radiation that has re-ionised the Universe. Since 
the direct detection of ionising radiation from the epoch of re-ionisation (EoR) is not possible because of the cosmic 
opacity, the low-redshift ``analogs'' of the distant galaxies play a key role in understanding the mechanisms that allow 
the escape of ionising radiation from star-forming galaxies. 

The census of LyC-galaxies is growing fast, both in the nearby Universe  \citep[][and references therein]{izotov18}
and at high-redshitf, i.e. $z\approx 3.5$,  \citep[][]{vanz18, vanz16b, debarros16,shapley16,bian17}, and 
relevant progress has recently been made in a statistical
sense by analysing dozens of high redshift galaxies with dedicated HST imaging \citep[e.g.,][]{felce18,jure17} and 
deep spectroscopy \citep[e.g.,][]{steidel18,marchi18}. In particular, a positive correlation among LyC escape and \lya\ equivalent
width has been inferred, as well as an apparently higher {\em fesc} at fainter ultraviolet magnitudes, such that galaxies might
account to more than 50\% of the ionising budget at $z\sim3$ \citep[][]{steidel18}.
Spectral features like the profile of the escaping \lya\ line, the strength of the low-ionisation interstellar absorption lines 
tracing the covering fractions of neutral gas (e.g., \cii, \siii)
the line ratios tracing the ionisation$-$ or density$-$bounded conditions in the interstellar medium 
(like the O32 index, \oiiidoub\ / \oiidoub), hold the promise to provide useful diagnostics 
of the mechanisms that govern the escape of ionising radiation, although we currently do not yet know which properties 
provide necessary and/or sufficient conditions for this to happen 
\citep[e.g.,][]{schaerer16,verhamme17,izotov18,jaskot13,mckinney19,gaza18,JC18,reddy16,reddy18,steidel18,grazian17}.
\begin{figure}
\centering
\includegraphics[width=8.5cm]{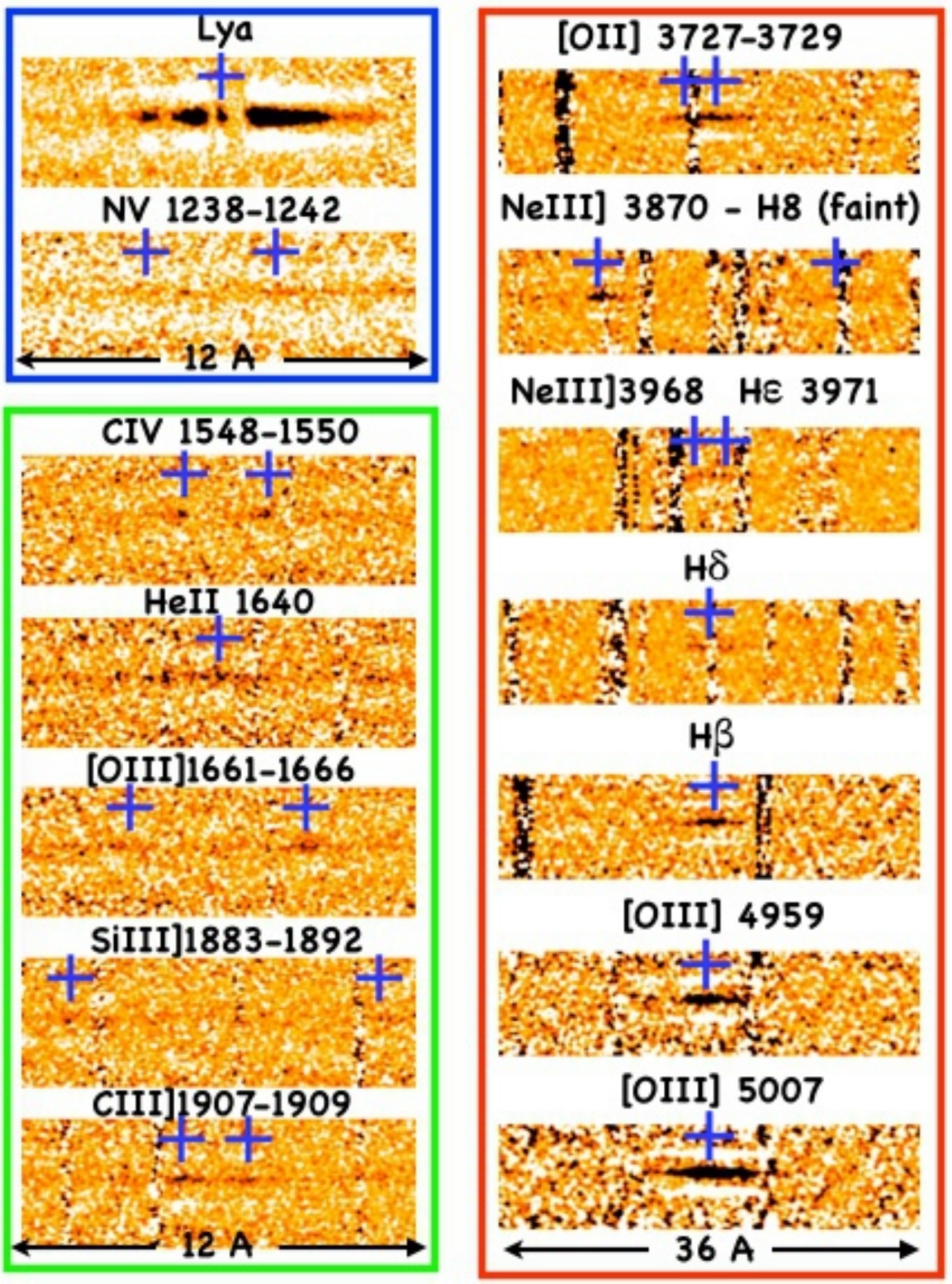} 
\caption{The most relevant atomic transitions of {\em Ion2} in the VLT/X-Shooter UVB (R=5300), VIS (R=8900) and NIR (R=5600) arms, 
colour-coded in blue, green and red boxes, respectively. The expected positions at systemic redshift z=3.2121 are indicated with
 blue symbols (`+'). For each transition, the corresponding rest-frame wavelength is reported. The rest-frame wavelength width of the 
zoomed spectra is reported at the bottom, for each arm. The significance of the reported transitions is summarised in Table~\ref{tab}.}
\label{lines}
\end{figure}
While the current growing samples serve as a reference for the identification of cosmic reionisers,
the physical processes that made these galaxies transparent to LyC radiation are yet to be understood.
The way ionised channels are carved in the interstellar medium is
under continuous investigation, especially in the local Universe
 \citep{herenz17,micheva17,micheva19,bik18,kehrig18} 
where the detection of LyC radiation is instrumental to catch such episodes in the act  \citep[e.g.,][]{h11}.
It remains unclear what is the spatial distribution of the escaping ionising radiation,
the porosity and the kinematics of the neutral gas, and the role the internal constituents of 
high-z LyC$-$galaxies (namely, the star-forming complexes, OB-associations, young massive 
clusters and massive stars) have in carving such ionised regions. 
In general, the small physical scales, likely of the order of 
a few ten pc and in which such constituents originate, are still unreachable at cosmological distances 
(e.g., a typical HST pixel of 30 milli-arcsecond subtends 250-300 pc at $z\sim 2-6$, 
encompassing one or more star-forming complexes). 
This limitation is even worse when ground-based seeing-limited spectroscopy
is performed. 

This work presents new VLT/X-Shooter spectroscopy of a LyC$-$galaxy dubbed {\em Ion2}
\citep[][]{vanz16b}, and VLT/MUSE observation of the recently discovered strongly lensed LyC emitter at z=2.37 \citep[known as Sunburst arc, ][]{dahle16,rivera17,rivera19}.
The similarity between the exceptional strongly$-$lensed Sunburst and {\em Ion2} sheds light on the possible ``engine'' behind the spatially
unresolved high-z LyC leakers, like {\em Ion2}. We make use of archival VLT/MUSE and HST/ACS data targeting the Sunburst 
 object with the aim to emphasise and explore this connection. 

We assume a flat cosmology with $\Omega_{M}$= 0.3,
$\Omega_{\Lambda}$= 0.7 and $H_{0} = 70$ km s$^{-1}$ Mpc$^{-1}$.

\section{The Lyman continuum galaxy {\em Ion2}}

{\em Ion2} is a well known LyC emitter at z=3.2121 lying in the CDFS showing an escape fraction 
higher than 50\%\ \citep[][]{vanz15,vanz16b}.
Here we present new VLT/X-Shooter observations that improve (in terms of depth, spectral resolution and wavelength coverage)
our previous analysis \citep{debarros16}.

\subsection{X-Shooter observations}
{\em Ion2} was observed during November 2018 for a total integration time
of 5 hours under optimal seeing conditions, typically $0.5 - 0.7$ arcsec. The slit widths were
1.0$''$, 0.9$''$, 0.9$''$, corresponding to a spectral resolution of R=5400, 8900 and 5600 in 
the UVB, VIS and NIR arms, respectively (Prog. 0102.A-0391(A),  P.I. Vanzella). Given the
good seeing, the aforementioned  resolution values must be considered as lower limits.
The data reduction was carried on as described in several previous works 
(we refer the reader to \citealt[][]{vanz16a,vanz17b}), in which the AB-BA sky subtraction scheme
was implemented with single exposures of 915s, 946s and 900s on each of the three UVB, VIS and 
NIR arms, respectively. The target was dithered $2.4''$ along the slit.
The spectral range from the U to the K-band covers the rest-frame
wavelengths which include the \lya\ and the optical lines \oiiidoub. 
The continuum is barely detected, from which a dozen of emission lines emerge with S/N ratio
spanning the interval $2-50$ (see Figure~\ref{lines} and Table~\ref{tab}).
A careful analysis of the statistical significance of the spectral features 
is reported in the Appendix~\ref{SN}.

\subsection{Results}

The X-Shooter spectrum shows at least three features not observed with previous spectroscopy \citep[][]{debarros16}: 
a multi-peaked \lya\ profile, the presence of narrow ultraviolet high ionisation lines and the new detection of \oiidoub\ and \hb\ optical 
rest-frame lines. Below we summarise these new results.

\begin{figure}
\centering
\includegraphics[width=8.0cm]{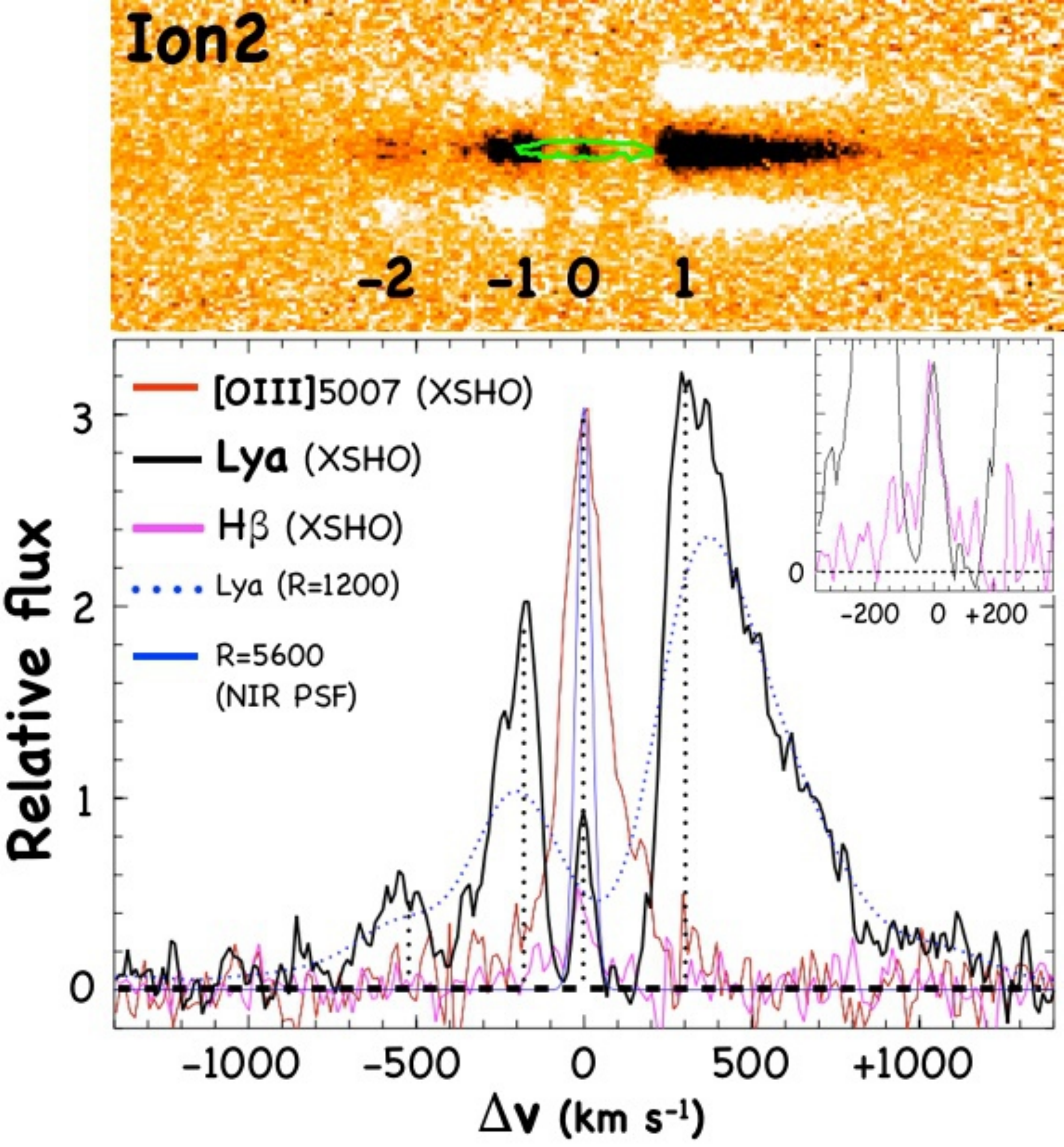} 
\caption{The top panel shows the two-dimensional \lya\ of {\em Ion2} (at resolution R=5300) with indicated the wavelength position 
of the \oiiiv\ (5$\sigma$ green contour) and the four peaks, labelled as `-2', `-1', `0' and `1'. The peak `0' falls exactly at the 
 \oiiiv\ redshift. In the bottom panel the \lya, \oiiiv\ (rescaled for clarity) 
and \hb\ line profiles are superposed in the velocity domain.
The inset shows the zoomed region around the zero velocity ($-400< \Delta v <+400$ \kms) where the \lya\ peak `0' has here
been rescaled to match the peak of the \hb\ line with the aim to emphasise the consistency among the widths of the lines.
The effect of low spectral resolution is also shown (R=1200, blue dotted line).
The Gaussian shape representing the spectral resolution in the NIR arm (R=5600) is also shown for comparison with a solid blue line. }
\label{ion2}
\end{figure}

\begin{figure}
\centering
\includegraphics[width=8.0cm]{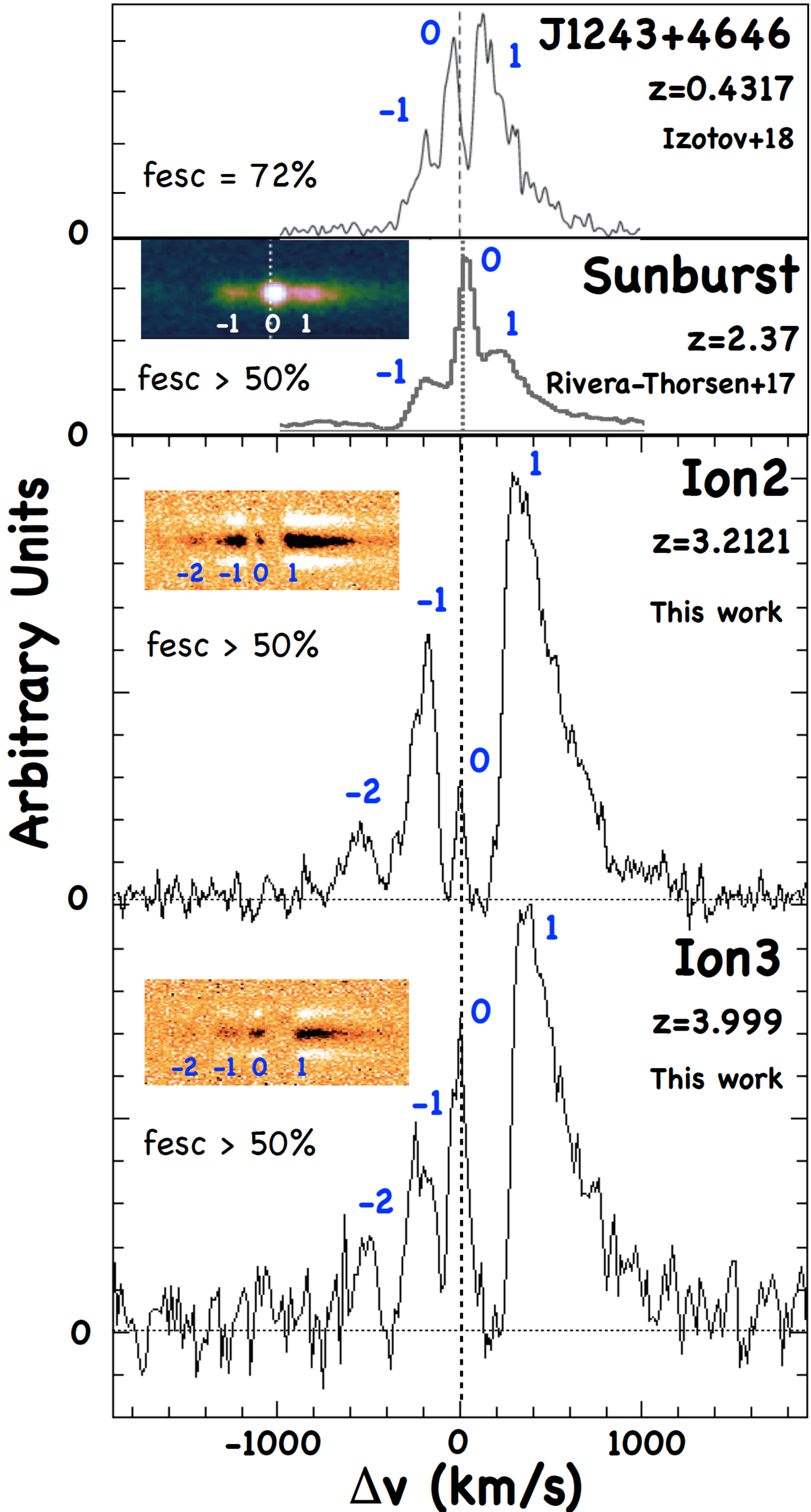} 
\caption{\lya\ profiles in the velocity space of all the LyC emitters with triple$-$peaked \lya\ emission
currently known. The upper two panels have been adapted 
from \citet{izotov18} and \citet{rivera17}. The insets show the two dimensional \lya\ spectra reported with their multiple peaks with
indicated the emission '0'  close to zero velocity (dotted line; see also Figure~\ref{ion2} for a detailed view in the case of {\em Ion2}). 
The relative escape faction ({\em fesc}) is also reported in each panel.} 
\label{lyas}
\end{figure}

\noindent $\bullet$ The \lya\ profile shows three peaks with the central one placed exactly at the systemic
redshift (z=3.2121, see Figure~\ref{ion2}), resembling the same \lya\ structure observed in another 
LyC$-$galaxy discovered at z=4.0 \citep[dubbed {\em Ion3,}][]{vanz18}. 
This is the fourth confirmed LyC emitter showing a \lya-peak emerging at the systemic velocity. The four objects are
also characterised by a quite large escape fraction of ionising radiation, {\em fesc} > 50\% (\citealt{vanz18,rivera19,izotov18}).
Figure~\ref{lyas} shows a comparison of the \lya\ shapes, and includes {\em Ion2}, {\em Ion3},
a high-surface density LyC emitter at z=0.4317 with a very large escape fraction ({\em fesc}~$>72$\%) from the \citet{izotov18} sample and Sunburst.
In the LyC emitters shown in Figure~\ref{lyas} the positions of the peaks at the two sides of the central one are different,
reflecting different kinematical features characterising each system.
The same line shape has been investigated by \citet{rivera17} performing radiative transfer (RT) calculations in the framework of three different scenarios 
including the expanding shell models \citep[][]{DS11,gronke15}:
a density-bounded medium, picket fence medium and the presence of a ionised channel embedded in the \hi\ shell.
Only the last case suitably reproduces the triple-peaked \lya\ profile: a significant amount of \hi\ gas with a perforated channel accounts for  both the typical
 \lya\ broadening by frequency diffusion (peaks far from the resonance frequency, e.g., `-2,-1,+1') and the superimposed \lya\ emission at zero velocity (peak `0').
Additionally, as discussed by \citet{rivera17}, the profile of the non-scattered \lya\ photons escaping through an optically thin tunnel would resemble the
width of the Balmer emission lines, that would represent a proxy of the intrinsic \lya\ shape before undergoing any RT effect. The brightest Balmer emission we have in the spectrum is the \hb\ line detected with S/N=6.
Figure~\ref{ion2} superimposes the central \lya\ peak (`0') and the \hb, that show compatible widths (being both marginally resolved, see Table~\ref{tab}). 
This is fully in line with what was predicted by \citet{behrens14} (see their Figure 7).
It is also worth stressing that the detection of the above narrow \lya\ features has been possible only  thanks to the high spectral resolution 
($R>5000$) achievable with X-Shooter (see the case R=1200 in Figure~\ref{ion2}), underlying the fact that the \lya\ line can be a powerful probe 
of optically thin media up to z=4 (and possibly up to z=6.5 in the case of transparent IGM,  e.g., \citealt{matthee18}).

\begin{figure*}
\centering
\includegraphics[width=17.5cm]{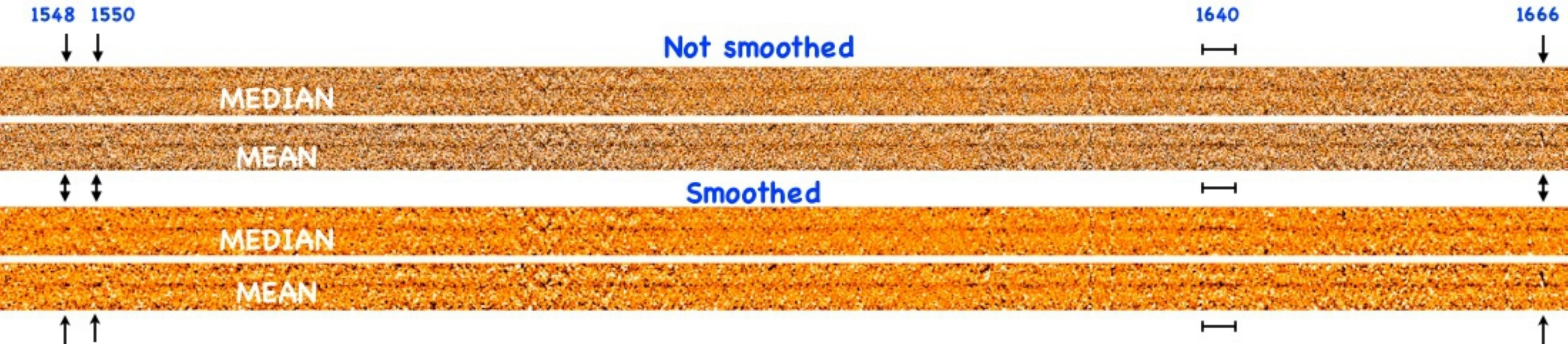} 
\caption{The two-dimensional X-Shooter spectrum of {\em Ion2} reporting the faint continuum and the \civ, \heii\ and \oiiiuv\ emission lines
is shown (indicated with arrows and a segment for the \heii\ emission). This figure shows the different broadness of the lines: 
the width of the segment is $\sim 500$ \kms, whereas the other \civ\ and \oiiiuv\ lines are much narrower and possibly not resolved ($\sigma_v < 25$ \kms). 
The top panels show the non-smoothed mean and median spectra. The same is reported in the bottom panels after applying a 
Gaussian smoothing with $\sigma = 1$ pixel.}
\label{HEII}
\end{figure*}

\noindent $\bullet$ The \civ, \oiiiuv, \heii\ and the \ciiidoub\ high-ionisation emission lines have been detected 
(see Figure~\ref{lines} and Table~\ref{tab}), with the doublets being well separated.
It is worth noting that all the nebular metal lines appear very narrow and possibly not 
resolved  ($\sigma_v < 20$ \kms), whilst the \heii\ emission, despite a relatively low S/N,
is clearly broader.  To emphasise such a difference, Figure~\ref{HEII} shows the portion of the X-Shooter two-dimensional spectrum
containing the \civ\ doublet, the \heii\ and \oiiiuv\ emission lines.  
The S/N of the \heii\ (3.5) is sufficient to appreciate its broadness, plausibly encompassing a velocity interval up to 500 \kms, marked in
Figure~\ref{HEII} with a segment (see also appendix~\ref{SN} for more details). 
It is worth noting that in other cases  the \heii\ emission is as narrow as the other high-ionisation metal lines \citep[e.g.,][]{vanz16a,vanz17c}
in which the nebular origin dominates or is better captured and other cases in which both nebular and broader stellar components
are measured \citep[][]{erb10,senchyna17,senchyna19}. In the case of {\em Ion2} the relative contribution of nebular and stellar
components is not measurable.  However, the presence of a broad \heii\ emission profile suggests a spectrum dominated by
a young stellar population containing hot Wolf-Rayet stars with main-sequence lifetimes less than 5 Myr \citep[][]{JC19}.

\noindent $\bullet$ Differently from the previous analysis based on a much shallower Keck/MOSFIRE spectrum in which
the optical \oiidoub\ and \hb\ lines were not detected  \citep[][]{debarros16}, 
here we measure a rest-frame equivalent width (EW) of EW(\hb) $\simeq 100$\AA, O32 $= 9.18_{-1.32}^{+1.82}$ and 
\oiiiv / \hb = $8.55_{-1.41}^{+1.96}$ (see Table~\ref{tab}). 
 Such O32 value is in line with the {\em necessary} condition of having a large O32 index in LyC leakers \citep{jaskot13,izotov18}.
It is worth noting that the rest-frame EW 
of \oiiidoub\ is 1300\AA, not dissimilar from the strong oxygen emitters found at $z>7$ \citep[e.g.,][]{castellano17,RB16}.  
This also suggests a relatively large ionising photon production efficiency ($\xi_{ion}$), defined as the production rate of \honly-ionising photons 
per unit intrinsic monochromatic UV luminosity ($\xi_{ion} \simeq 25.6$, following \citealt{chevallard18}).\footnote{Note that
the \ha\ line is not accessible and the \hb\ is detected at low S/N and affected by dust extinction and possibly damped by the LyC leakage.}

While the \lya\ profile further  confirms {\em Ion2} to be a genuine LyC emitter,
the detailed geometry of the LyC emission and the origin of the ionising radiation is still
unknown \citep{vanz16b}. 
Not surprisingly, if the
ionised channel and/or the size of the source emitting LyC radiation (namely, the region including O-type stars) 
is confined within a few tens pc or less (see Sect.~\ref{ion2asymc}), the WFC3/F336W spatial resolution
would be insufficient to resolve the source (1 pix $\simeq 150$ pc). Any further detailed investigation in the
rest-frame ultraviolet/optical bands would therefore be postponed to future studies with larger telescopes.
Before the advent of E$-$ELT-like telescopes that will provide a spatial resolution lower than 10 mas 
(corresponding to $\sim $ 75 pc at the redshift of {\em Ion2}), the only way to address individual star-forming complexes 
of a few tens pc require strong gravitational lensing \citep[e.g.,][]{vanz19,vanz17c,cava18,rigby17,jonhson17}. 
Even more valuable would be the identification of strongly lensed 
galaxies showing escaping LyC radiation emerging from some of their internal constituents. 
This happened recently with the discovery of the {\em Sunbust} arc and is the argument of the next section.

Before discussing it, it is worth stressing that the requirement of having the simultaneous alignment
of the ionised channel, the observer, the presence of shot-lived O-type stars and the transparent IGM along the 
line of sight implies that the visibility of the LyC radiation from high redshift sources 
is affected by severe view-angle and l.o.s. effects (e.g., \citealt{terlevich17,cen15,wise14}), 
not to mention the insidious foreground contamination mimicking false LyC radiation 
\citep[e.g.,][]{vanz10,vanz12,siana15}. 
Altogether, these effects make the detection of LyC$-$galaxies at high 
redshift still elusive and suggest that a significant fraction of them might be hidden by the aforementioned effects. Moreover, if we require that the source is also strongly magnified by an intervening gravitational
lens, then the event would be extremely 
rare.\footnote{The lensing cross section for events with magnification $\mu$ exceeding the 
threshold $\mu_0$ decreases rapidly with the square of the magnification itself: $\sigma_{lens} (> \mu_0) \sim \mu_{0}^{-2}$.}
The identification of a few LyC$-$galaxies either in non lensed or lensed fields
therefore makes the current detections extremely precious,
especially if we focus on the spectral similarities among these uncorrelated objects.

\section{Discussion}

To shed more light on the nature of {\em Ion2}, key information might be extracted by comparing 
its X-Shooter spectrum to the strongly magnified LyC emitter at z=2.37, dubbed Sunburst arc \citep[][]{dahle16,rivera17,rivera19}.
In this work we highlight the similarities between such systems arguing that what is observed in Sunburst is
compatible with what is currently hidden by the limited spatial resolution.

\begin{figure*}
\centering
\includegraphics[width=17.5cm]{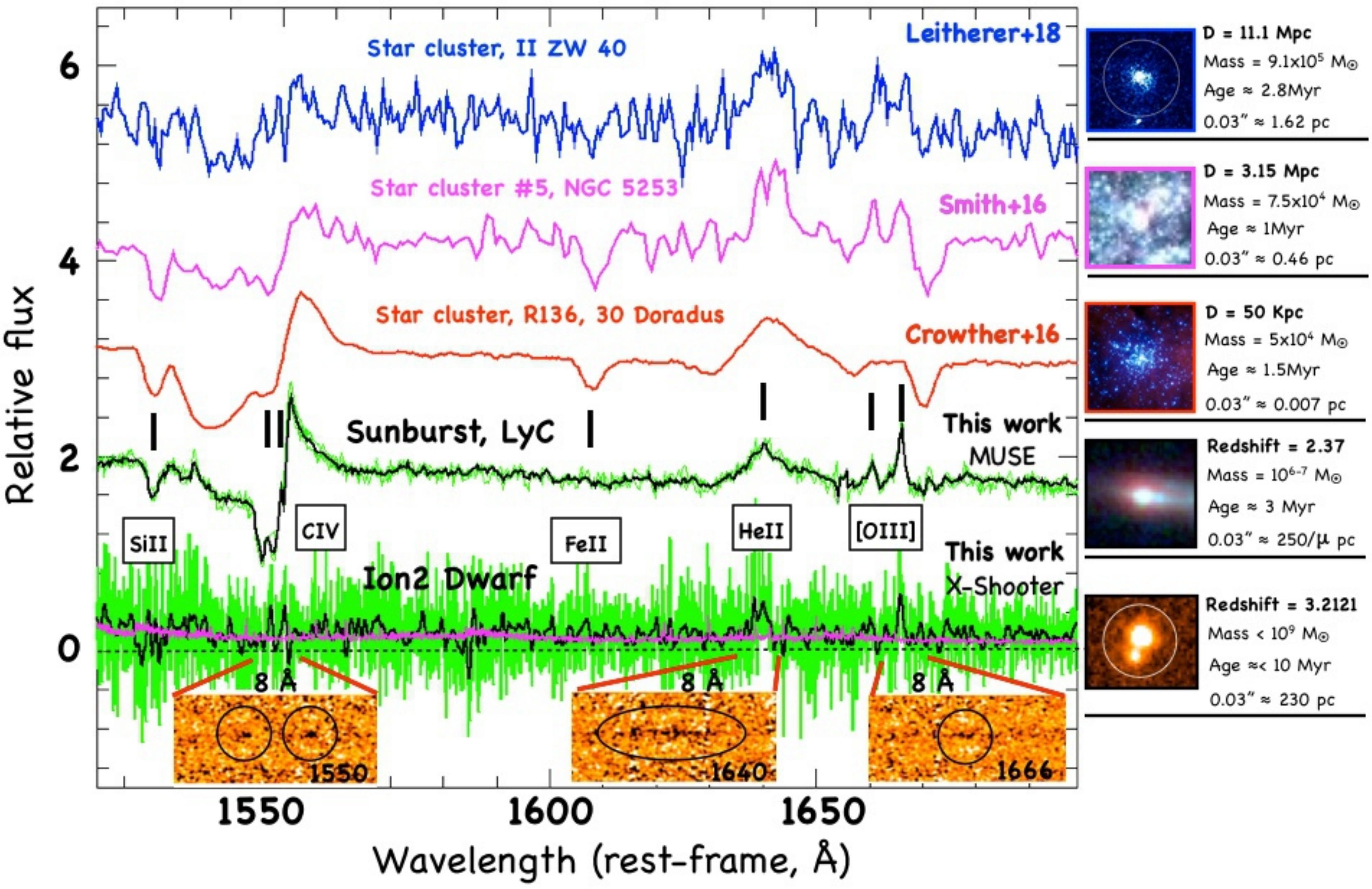} 
\caption{The ultraviolet spectra of local super star clusters (SSCs), the Sunburst
and the {\em Ion2} are shown. 
The strongly-lensed Sunburst MUSE spectrum (black line as the sum of 7 multiple images, plotted
with  green lines) is remarkably similar to local SSCs. The spectra of {\em Ion2} at the original spectral 
resolution (green) and smoothed at the MUSE resolution (black) show nebular metal lines with a 
broad \heii\ emission, as visible from the three insets at the bottom. On the right side,
some basic information on each system is reported, including distance (in Mpc), stellar 
mass (\msun), age (in Myr) and size (in arcseconds and pc). 
In the case of Sunburst, the size is a function of the magnification $\mu$, estimated to be of the order of $50-100$ \citep{dahle16}.
}
\label{clusters}
\end{figure*}

\subsection{The Sunburst arc} 

Sunburst is an exceptionally bright (R-band magnitude $\sim 18$) gravitationally lensed arc in which
the galaxy cluster PSZ1-G311 produces a magnification of the order of $\sim 50$ or even larger \citep[][]{dahle16}.
Specifically, four multiple arcs are generated by the galaxy cluster, and in the most magnified ones
additional amplification is generated by individual galaxy cluster members.
One of the star-forming knots of the arcs has also been detected in the LyC (hereafter dubbed LyC-knot) 
with the unprecedented record of 12 multiple images \citep[][]{rivera19} with a measured
magnitude in the interval F814W = [$21.5 - 22$] and detected with S/N$>30$ for the most magnified 
ones.\footnote{The magnitudes reported by \citet{rivera19} are about half a magnitude fainter than our estimates,
while the magnitude contrast among the multiple images is fully consistent with their estimates.
As shown in the appendix~\ref{minmag} this difference is mainly due to the small aperture they used ($0.12''$ diameter),
motivated by the need to measure the ratio between the ionising and non-ionising fluxes of a non-spatially resolved source.}

We present for the first time VLT/MUSE
observations targeting the Sunburst and based on the DDT programme 297.A-5012(A) (PI. Aghanim).
The data were acquired during May-August 2016 with seeing $0.5''-0.8''$ for a total integration of 1.2h (three exposures of 1483s each). 
The data reduction has been performed as described in \citet{caminha17,caminha19}, and we refer the reader to those
works for details.

A stacked MUSE spectrum obtained from seven multiple images
is presented in Figure~\ref{clusters}. In the same figure, some high-quality spectra of  local star clusters 
are shown for comparison (for further details, see the caption of Figure~\ref{clusters}). 
The ultraviolet emission of the LyC$-$knot resembles those of local super star clusters in which
the signatures of massive stars are clearly imprinted in the spectrum, like the prominent P-Cygni profiles of the \civ\ doublet and 
a broad \heii\ ascribed to the presence of Wolf-Rayet stars. The similarity with the ultraviolet spectra of a few 
well studied local young massive clusters is remarkable, namely 
R136 \citep[][]{crowther16}, II~Z40 \citep[][]{leitherer18} and 
cluster \#5 of NG5253 (\citealt[][]{smith16}, see also \citealt{calzetti15}), as well as the analogy with the nearby 
star forming regions collected by \citet{senchyna17,senchyna19} and showing high ionisation metal lines 
and in some cases broad \heii\ emission of low metallicity massive stars.
The same spectral features, though at slightly lower S/N ratios, have been identified and accurately modelled 
by \citet{JC19}, providing a stellar age of $3.0\pm0.1$ Myr and subsolar stellar metallicity Z = $0.60\pm0.05$ Z$_{\odot}$, with
an inferred dust extinction E(B-V) $\simeq$ 0.15.

While these are very precious quantities and independent from the lensing magnification, it is now natural to try addressing
intrinsic quantities such as the stellar mass, the luminosity and the physical size of such LyC emitter. This will be the argument of the
next section.

\subsection{The Sunburst LyC-knot as a possible gravitationally bound system}

We express the most relevant physical quantities, namely the effective radius, the stellar mass, and the 
stellar mass surface density ($\Sigma^{\star}$), as a function of magnification $\mu_{TOT}$, 
adopting $R_e =  2.0$ pix, e.g., 60 mas along the tangential direction as derived by performing {\em GALFIT} fitting
(see Appendix~\ref{minmag} and Figure~\ref{profiles}).
The stellar mass is estimated assuming an instantaneous burst with the aforementioned age of 3.0 Myr and
sub-solar metallicity Z~=~0.6~Z$_{\odot}$, 
and adopting {\em Starburst99} models \citep{leitherer14} with a Salpeter initial stellar mass function (IMF, $\alpha = 2.3$), 
including stars with masses in the range [1-100] \msun.
Similarly, we derive the stellar mass using a top-heavy initial stellar mass function with a slope $\alpha = 1.6$, following \citet[][]{tereza17}, 
with masses in the range [0.1-120] \msun. The two IMFs should embrace two extreme cases 
with the aim to provide a lower and an upper limit to the stellar mass.

\begin{figure}
\centering
\includegraphics[width=8.5cm]{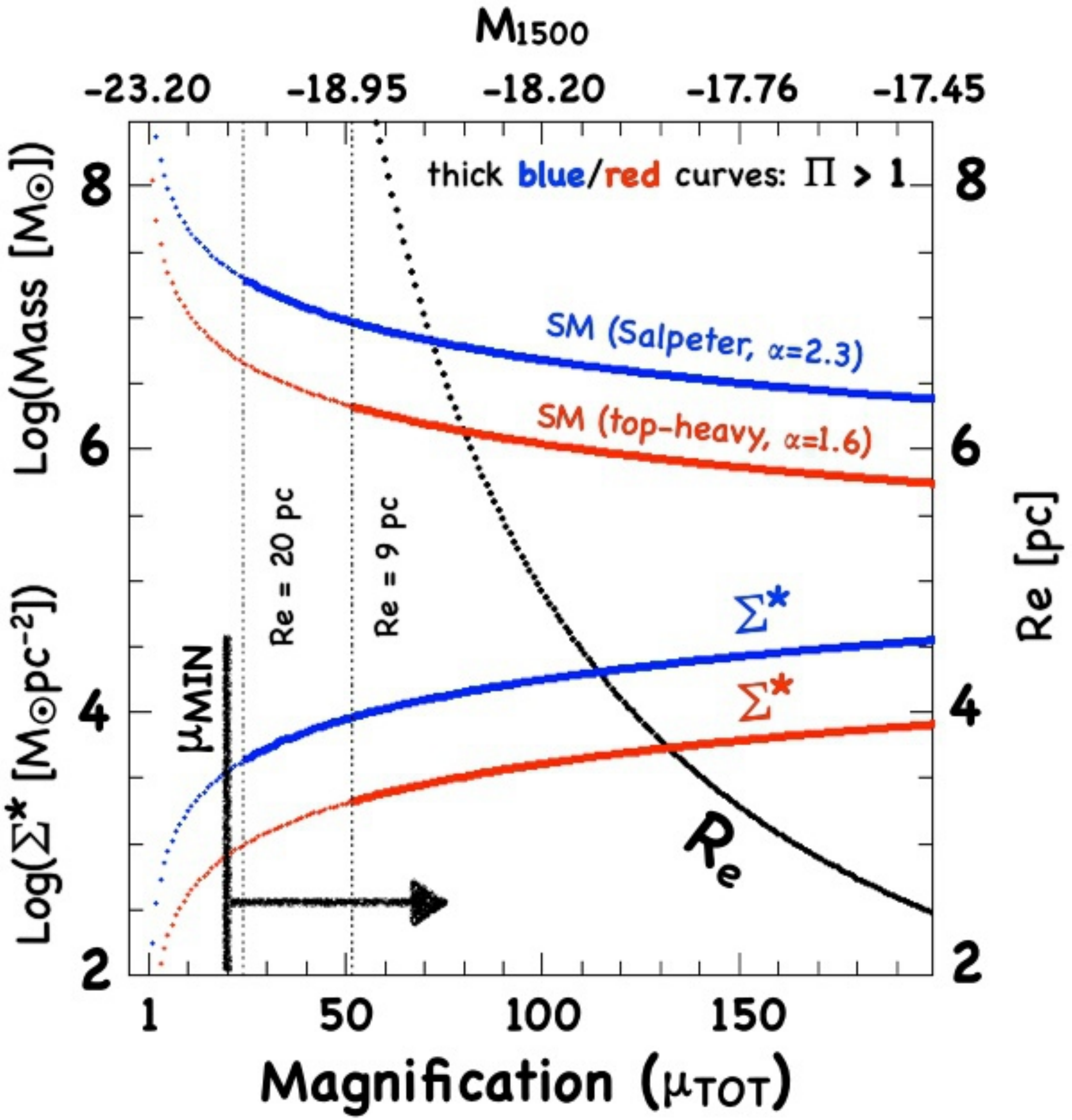} 
\caption{The stellar mass, stellar mass surface density ($\Sigma^{\star}$) and the effective radius ($R_e$, on the right Y-axis) of the Sunburst LyC-knot
are calculated as a function of the total magnification, $\mu_{TOT}$. Red and blue curves correspond to top-heavy and Salpeter IMF,
respectively. The LyC-knot becomes compatible with a 
gravitationally bound star cluster if $\mu_{TOT} > 25(50)$. This regime ($\Pi > 1$) is highlighted with the thick lines, where $\Sigma^{\star}$
approaches the values measured in the densest objects known, like globular clusters and young massive clusters. The thick black horizontal arrow
in the bottom marks the minimum magnification, $\mu_{TOT} > 20$, we estimate in appendix~\ref{minmag}.}
\label{mass}
\end{figure}

As shown in Figure~\ref{mass}, an effective radius ($R_e$) smaller than $20$ pc is found if $\mu_{TOT} > 25$, and it decreases below
9 pc if $\mu_{TOT} > 50$. The stellar mass ranges between $10^{6} - 10^{7}$ \msun\  depending on the IMF and magnification, with a $\Sigma^{\star}$ that
enters the regime of the densest objects known (e.g., the globular clusters, e.g., \citealt{hopkins10}) or
similar to the values of young massive star clusters \citep[][]{bastian06,ostlin07,bastian13}.

Combining $R_e$ with the age of 3.0 Myr, and the stellar mass, we can infer
the dynamical age  as $\Pi =$ Age/ T$_{cr}$ \citep[where T$_{cr}$ is the crossing 
time: 10($R_e^{3}$/GM)$^{0.5}$,][]{gieles11}. Interestingly, as highlighted in Figure~\ref{mass}, 
the system enters the regime of a gravitationally bound object ($\Pi > 1$)  if 
$\mu_{TOT}>25(50)$ in the case of Salpeter(top-heavy) IMF. 
Such magnification values are within the expected magnification 
regime \citep[e.g.,][]{dahle16}.  In particular, a {\it minimum} model-independent estimate of the magnification of 20
is derived and discussed in the appendix~\ref{minmag}, based on empirical geometrical constraints.
Therefore, the LyC$-$knot might be the first example of a gravitationally-bound
star cluster discovered at cosmological distance. 

\subsection{Is also the LyC$-$galaxy {\em Ion2}  powered by star clusters?}
\label{ion2asymc}

The  discovery of a very likely gravitationally bound star cluster at z=2.37 leaking 
LyC radiation is intriguing because it would imply that the contribution by such systems to the meta-galactic ionising 
background is substantial, if not dominant, depending on the UV luminosity function of such objects. 

Observationally, 
Sunburst$-$like objects in which the LyC leakage emerges from a single massive star cluster of a few pc cannot
be spatially resolved even in moderately lensed fields, e.g. $\mu < 20$. 
Compact SF$-$clumps at high redshift, either spatially resolved or not, may be dominated by single young massive star clusters  \citep[e.g.,][]{zanella15,jonhson17,rigby17}.

Any seeing-limited spectrum would be the luminosity-weighted average of multiple unresolved
star$-$forming complexes, as for the case of {\em Ion2} in which structures smaller than 200 pc cannot be resolved.
However, the similitude among the spectral properties of {\em Ion2} and Sunburst is intriguing and
might offer new clues, beyond the limitation due to the spatial resolution.
The ionised channels traced by the triple$-$peaked \lya\ profiles $-$ especially the narrow peak 
at systemic velocity $-$ of {\em Ion2}, {\em Ion3} and Sunburst (including the local system of \citealt{izotov18})
might suggest a common origin related to the presence of young massive star clusters
and/or dense star-forming regions. The HST imaging of {\em Ion2} shows a quite
nucleated morphology \citep[the LyC emission is spatially unresolved,][]{vanz16b}, 
as well as the triple$-$peaked \lya\ object of \citet[][]{izotov18} that shows the highest
star formation rate surface density ($> 500$ \msun\ yr$^{-1}$kpc$^{-2}$) in their sample.  
Currently, the Sunburst LyC-knot seems to be the densest stellar LyC leaker with also 
emergent \lya\ at systemic velocity. In addition, the broad \heii\ emission observed in {\em Ion2} 
(possibly with FWHM $> 400$ \kms, see appendix~\ref{scan}) and the 
well detected \nv\ P-Cygni profile of {\em Ion3} \citep[][]{vanz18} suggest  a radiation leakage through
one or more channels carved  by massive stars promoted by their energetic feedback
might be in place, as observed in the Sunburst.

\section{Conluding remarks}

In this paper we have presented new VLT/X-Shooter observations of the LyC emitting galaxy {\em Ion2}
and VLT/MUSE spectrum of a strongly lensed LyC emitter, dubbed Sunburst. The results can be summarised as follows:

\begin{itemize}
{\item {\em Ion2}: the spectral resolution and wavelength coverage provided by X-Shooter have improved
the previous analysis presented in \citet{debarros16}, including the detection of new spectral features. 
First, a multi-peaked \lya\ profile is evident from the new spectrum, showing a clear emission at the systemic redshift.
Second, several high ionisation ultraviolet nebular narrow lines (FWHM $<$ 50 \kms, including \civ, \oiiiuv, \ciiidoub) have been detected for the 
first time, some of which with well separated doublets. Only the \heii\ is detected as a broad emission (FWHM $>$ 400 \kms)
and ascribed to the presence of Wolf-Rayet stars. Third, a large value for the O32 index of $9.18_{-1.32}^{+1.82}$ has been derived, together with the
large equivalent width of 1300\AA~rest-frame of \oiiidoub, as found in other LyC leakers \citep{jaskot13,izotov18} and typical of 
systems with a large ionising photon production efficiency \citep[e.g.,][]{chevallard18}.}
{\item {\it Sunburst}: we inferred for the first time the stellar mass (a few $10^{6}$ \msun), luminosity ($M_{UV} > -19$) 
and size ($R_e < 20$ pc) of the LyC-knot of the Sunburst arc,
that coupled with the young age of 3 Myr \citep{JC19}  provides constraints on its dynamical age,
 suggesting that the LyC-knot is a gravitationally$-$bound young massive stellar cluster at
 cosmological distance, whose ultraviolet spectrum is also fully comparable to those of local young clusters 
 (Figure~\ref{clusters}). In addition, as the Figure~\ref{mass} shows, the stellar mass surface density is intriguingly large
 if the magnification factor exceeds 50, approaching the values observed in the densest objects, such as globular clusters and local young massive clusters \citep[][]{hopkins10,bastian06,ostlin07,bastian13}. Remarkably,  Sunburst might also be considered a forming globular cluster, caught when the Universe was 2.7 Gyr old. This will be investigated in a future work.}

\end{itemize}

The LyC-knot of the Sunburst arc might very well represent the Rosetta stone of stellar ionisation at 
high redshift and it is an unprecedented discovery in its own right for two reasons:
(1) it is a unique laboratory where the escaping LyC from a high redshift stellar system can be investigated in detail 
and (2) without any lensing effect, the LyC-knot (the star cluster) would have appeared like a non-spatially resolved
LyC$-$emitter lying somewhere within its hosting galaxy, e.g., like {\em Ion2}.  

Interesting enough, the star cluster formation efficiency, namely the star formation occurring in gravitationally bound star clusters,
increases with redshift \citep[e.g.,][]{pfeffer18}, possibly reaching values higher than $30$\% at $z>6$. This might suggest that 
star clusters could have played a significant role during
reionisation \citep[][]{ricotti02,ricotti16}, especially if the LyC leakage is more efficient for that population. 
The idea that such young massive star clusters (or a fraction of them) 
were also globular cluster precursors is currently matter of investigation 
(e.g., \citealt{vanz19,renzini17,pozzetti19,bouwens17,pfeffer18,elme18,kruijssen19,RC19,calura15,calura19,li19}).

While the direct detection of LyC radiation at $z>3$ is challenging and
requires that special conditions are realised in the source, with current sensitivity, the effect 
that the transverse LyC leakage has on the surrounding medium might be easily detectable. 
Objects like {\em Ion2/Ion3} or Sunburst having transverse leakage of LyC radiation could induce 
spatially offset \lya\ or Balmer series fluorescence \citep[e.g.,][]{mas-ribas17}. 
Spatially offset \lya\ emission/nebulae
routinely detected with integral field spectrographs (like MUSE) might represent a viable tool to search for
possible local escaping ionising radiation around star-forming galaxies  \citep[e.g.,][]{vanz17a,vanz17c,wisotzki18,gallego18}.

Finally, the prospects for future investigations
of star-formation at very small scales $-$ down to single star clusters $-$ at cosmological distance appear very promising. 
In particular objects like Sunburst LyC-knot  stretched by magnification factors
larger than 30 will be probed down to FWHM(or pixel-scale) of 6(2) and 3(0.4) pc by VLT/MAVIS and ELT MAORY-MICADO, respectively.
These two MCAO$-$assisted\footnote{MCAO = Multi Conjugate Adaptive Optics.} instruments will be also complementary in 
terms of wavelength coverage, probing the ultraviolet and optical rest-frame wavelengths. Integral field spectroscopy at VLT or ELT
will also probe the spatial distribution of nebular high ionisation lines, as a signature of possible stellar mass segregation in star complexes
and providing maps at pc scale opening for two-dimensional studies of feedback 
mechanisms and star formation processes \citep[e.g.,][]{james16},
at cosmological distances.

\begin{table}
\footnotesize
\caption{The most relevant atomic transitions of {\em Ion2} are reported, showing the corresponding 
zoomed regions on the two-dimensional spectrum of Figure~\ref{lines}.
The S/N ratios indicate the reliability of the lines (see also appendix~\ref{SN}). 1-$\sigma$ upper limits on the
 line fluxes are reported in the case of non detections. The \nev, \hg\ and the \auroral\ are not
reported as they lie on the atmospheric absorption bands.
Line fluxes are reported in units of $10^{-17}$ \ergscm\ (no slit losses are considered) and the FWHM is expressed in \kms; the rest-frame equivalent width (EW) is reported in \AA\ and calculated starting from the line fluxes and continuum derived from CANDELS photometry, adopting magnitudes $\simeq 24.45$ and $\simeq 24.1$ in the optical and near infrared arms, respectively.
The comment ``Narrow'' means the line is not resolved, while ``Broad'' stands for resolved but a precise measurement is not 
feasible.}
\begin{tabular}{l l l} 
Line/$\lambda_{vacuum}$ & Flux($\frac{S}{N}$)(FWHM)(EW) & Redshift($1\sigma$)\\ 
\hline
\lya(-2) 1215.7           &   1.68(8.0)($\sim$141)(5.7) & 3.2044(3) \\ 
\lya(-1) 1215.7           &   5.28(29.0)(146)(18.0) & 3.2096(1) \\ 
\lya(0) 1215.7           &   0.96(10.5)($<56$)(3.3) & 3.2121(2) \\ 
\lya(1) 1215.7           &   20.25(110)(298)(69.1) & 3.2164(1) \\ 
\lya(total)                  &   28.16($-$)($-$)(97.6)   & \\          
\nvalone\ $\lambda 1238.82 $  &  $<$0.12(1.0)($-$)($<0.5$) & (3.2121)fixed \\
\nvalone\ $\lambda 1242.80$  &  $<$0.12(1.0)($-$)($<0.5$) & (3.2121)fixed \\
\civalone\ $\lambda 1548.20 $  &  0.28(5.2)(Narrow)(1.5) & (3.2121)fixed \\
\civalone\ $\lambda 1550.78 $  &  0.20(4.8)(Narrow)(1.1) & (3.2121)fixed \\
\heiialone\ $\lambda 1640.42$  &  0.45(5.0)(Broad)(2.8) & (3.2121)fixed \\ 
\oiii\ $\lambda 1660.81          $  &  0.40(3.5)(Narrow)(2.5) & (3.2121)fixed \\
\oiii\ $\lambda 1666.15          $  &  0.65(5.4)(Narrow)(4.2)      & (3.2121)fixed \\ 
\siiiialone\ $\lambda  1882.65 $  & 0.30(2.0)(Narrow)(2.4)     & (3.2121)fixed \\
\siiiialone\ $\lambda  1892.03 $  & 0.15(1.5)(Narrow)(1.2)   & (3.2121)fixed \\
\ciiialone\ $\lambda  1906.68 $  & 0.44(6.7)(Narrow)(4.1)       & 3.2127(5) \\ 
\ciiialone\ $\lambda  1908.73 $  & 0.35(4.9)(Narrow)(3.2)     & (3.2121)fixed\\
\oii\ $\lambda 3727-3729$     &   3.5(6.0)($-$)(81) & 3.2122(3) \\ 
NeIII] $\lambda 3869.81$     &     1.8(4.0)($-$)(45)     & 3.2119(4) \\ 
NeIII] $\lambda 3968.53$     &    1.2(1.5)($-$)(32)     & (3.2121)fixed \\ 
H$\delta$ $\lambda 4102.92$ &    1.1(2.0)($-$)(31)     & (3.2121)fixed \\     
\hb\ $\lambda 4862.69$     &       2.9(6.0)($<60$)(114)    & 3.2120(3) \\ 
\oiiialone\ $\lambda 4960.30$     &    7.4(22.0)($-$)(304) & 3.2121(2) \\ 
\oiiialone\ $\lambda 5008.24$     &    24.8(65.0)(146)(1040)    & 3.2121(1) \\ 
\hline
\hline 
\end{tabular}
\label{tab}
\end{table}

\section*{Acknowledgments}
We thank the referees for providing detailed comments and suggestions about  the 
reorganisation of the manuscript.
We thank C. Leitherer, L. Smith and P. Crowther for providing the ultraviolet
spectra of the star clusters shown in Figure~\ref{clusters}. 
We thank A. Renzini, A. Adamo for stimulating discussions about the possible origin of the
LyC-knot.
EV also thank J. Chisholm and E. Rivera-Thorsen for stimulating discussions about the Sunburst system. 
EV thanks T. Jerabkova for fruitful interaction about the IMF used in this work.
EV also thank M. Gronke for useful discussions about the  \lya\ profile.
AM acknowledges funding from the INAF PRIN-SKA 2017 programme 1.05.01.88.04.
KC acknowledges funding from the European Research Council through the award of the Consolidator Grant ID 681627-BUILDUP.
We also acknowledge funding from the INAF for ``interventi aggiuntivi a sostegno della ricerca di main-stream dell'INAF''.

\appendix
\section{An empirical minimum magnification for the Sunburst LyC-knot}
\label{minmag}

\begin{figure*}
\centering
\includegraphics[width=17.5cm]{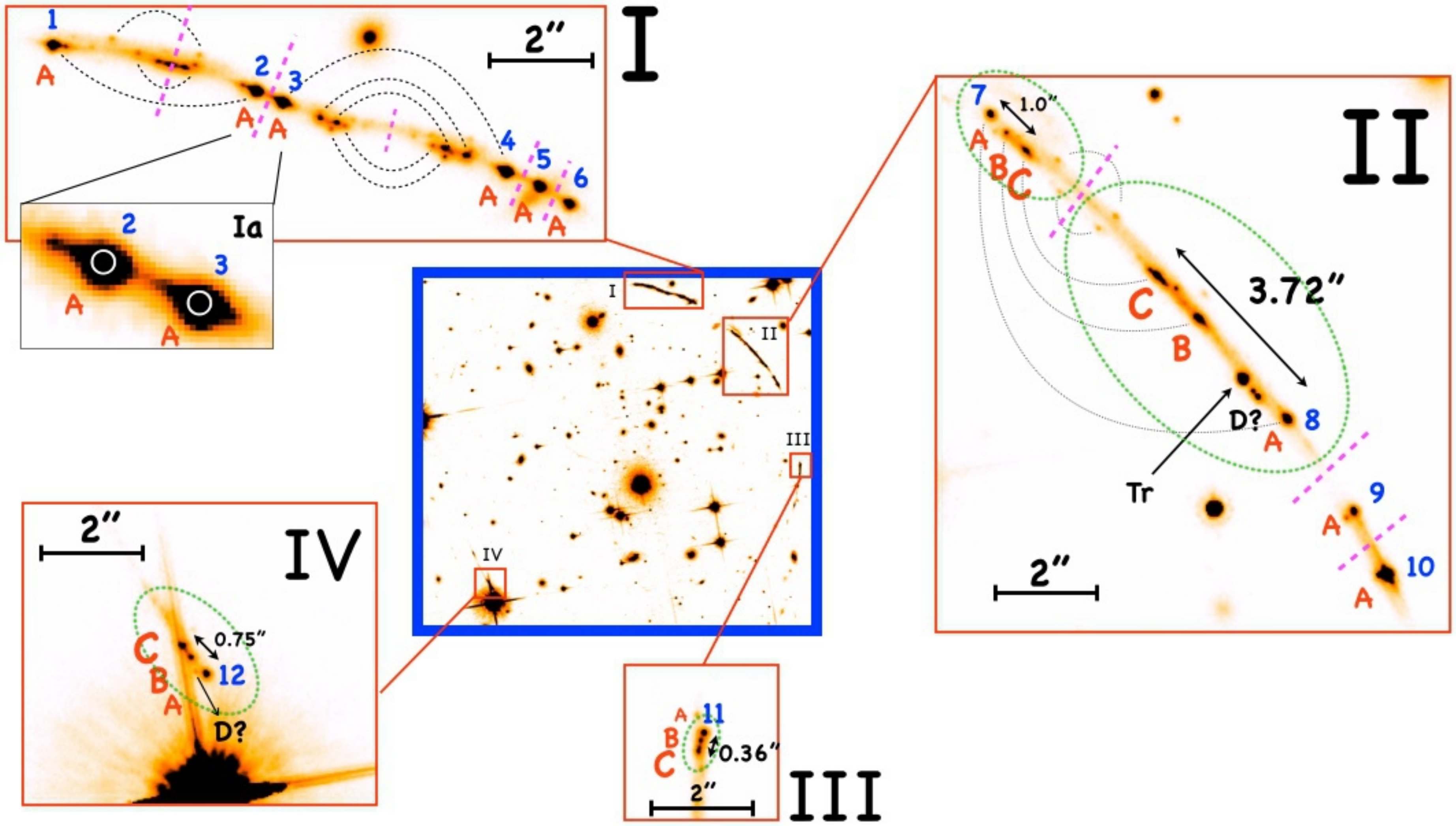} 
\caption{A complete overview of the four arcs are shown in the central panel with the zoomed regions
shown clockwise (panels I, II, III and IV). The tangential stretch is clearly visible after the identification of the
three knots A, B and C. The knot A is the LyC-knot discussed in this work and its distinctive signature allows a secure identification \citep[][]{rivera19}. 
In particular, the ratio among the angular separations between A-C for the less magnified system (III) and the most stretched one (II) implies an average 
tangential magnification larger than 10 for system II (i.e., knowing the arc III has $\mu > 1$). 
The relative flux ratio among the LyC-knots 2 (or 3) and image 8 of $\simeq 2$ further implies the
magnification for images 2 (o 3) is higher than 20. The magenta dotted lines mark approximately the locations where the critical lines cross the arcs
(for the arc I see also \citealt{rivera19}). 
The dotted green ellipses highlight the multiple structures containing the knots A,B,C and other small features. 
Some of the mirrored multiple images within the arcs (accordingly with the parity of the images genrated by strong lensing, \citealt{kneib11}) 
and tiny SF knots are connected with dotted black lines.
Other sub-structures are marked with `D'  and are visible only in the most magnified regions (or represent foreground sources). Finally, another bright
object not present in any of other arc is marked as `Tr' in panel II and represents a possible transient 
(Vanzella et al. in preparation). The inset Ia shows a zoom of the images 2 and 3 and highlights the circular apertures used in
\citet{rivera19} to derive the F814W magnitudes reported in their Table~1. While such a small aperture is appropriate for their
goals (measure of the ionising to non-ionising flux ratio for a point-like object), our magnitudes derived with {\em Galfit} fitting in the F814W band account for the total 
flux and spatially resolved morphology, making our estimates closer to the total magnitude.}
\label{fullsystem}
\end{figure*}

The sub-structures and star-forming knots  present in the Sunburst arcs allow us to set empirical constraints on the minimum (average) 
magnification among the recognised multiple patterns. Figure~\ref{fullsystem} shows the four arcs labeled I, II, II and IV, within which 
12 multiple images of the LyC-knot (indicated with `A') have been discovered by \citet{rivera19} and are marked with increasing numbers 1-12. In particular
the least magnified arcs provide the less distorted version of the high-z SF complex. Starting from the smallest arc III the knots A, B, C are
identified in the counter arc IV, with the inverse order (mirroring) of the SF knots as expected in strongly lensed multiple images. 
In this case the identification is further facilitated by the fact that the knot A has the distinctive LyC feature and the image C is the most elongated among
the three objects. Such features must follow the aforementiond flipped behaviour, accordingly to the parity of the images introduced by the strong
lensing \citep[][]{kneib11}.
Other two triplets A, B, C appear in the arc II where the mirrored
groups are circled with green dotted ellipses. The triplet separated by $3.72''$ is 10 times more elongated than the triplet in the arc III, that means the
average tangential magnification for the widest arc is larger than 10, being the magnification of arc III larger than 1. 

The situation of arc I is different because not all images (A,B,C) are present, i.e. only a part of the source is multiply lensed in this merging arc. However, 
six multiple images of knot A (the LyC emitter) are clearly identified (images 1-6 of Figure~\ref{fullsystem}). The proximity of some of the images to the
critical lines, like images 2 and 3, suggest the magnification is large \citep[see, e.g., Figure 4 of ][]{rivera19}. 
In particular the measured flux ratio between images 2  (or equivalently 3) and 8
is $\simeq 2$ \citep[][]{rivera19}, suggesting images 2,3 have a minimum magnification of $\sim20$ ($2 \times 10$). This happens for the brightest image 10, that is 2.2 times brighter than image 8. Until this point we did not invoke any lens model. The true magnification is higher since we are assuming
arc IV is not magnified ($\mu=1$). However, arc IV is subjected by strong lensing as well (being one of the multiple images of the system) 
and its magnification is certainly higher than 1.
For example, assuming a magnification 2 for arc IV, following the aforementioned considerations the magnification for images 2,3 jumps to values of 40.
This provides an average magnification, and as discussed above the proximity to the critical lines
suggests the lensing amplification at the position of, e.g., images 2-3, could be be much higher, even above 50
as discussed by \citet{dahle16}. A careful modelling of the lens, including the aforementioned empirical constraints, will shed more light
on this.

However, the minimum magnification of 20 derived above is already relevant in our study, implying an effective radius smaller than 20 pc and a stellar
mass in the range $5\times10^{6} - 2.5 \times 10^{7}$ \msun, depending on the IMF (see Figure~\ref{mass}), with the system entering the
range of massive, gravitationally-bound, star clusters in the case of Salpeter IMF.

It is worth noting the presence of a bright and point-like object with F814W = 22.02 and at the same redshift of the Sunburst 
(marked with `Tr' and an arrow in Figure~\ref{fullsystem}). Such an object is presumably a transient for two reasons: 
(1) it is not identified in any of the other arcs and (2) it shows unique spectral properties in the MUSE spectrum not observed in any of 
the other knots populating the rest of the arcs (the ongoing X-Shooter programme\footnote{VLT/X-Shooter, P103.A-0688(A-C), PI Vanzella.} will investigate `Tr' and the results presented elsewhere). 
The characterisation of such a transient (like the absolute magnitude), will provide a unique constraint for the
lens model.

\begin{figure}
\centering
\includegraphics[width=8.5cm]{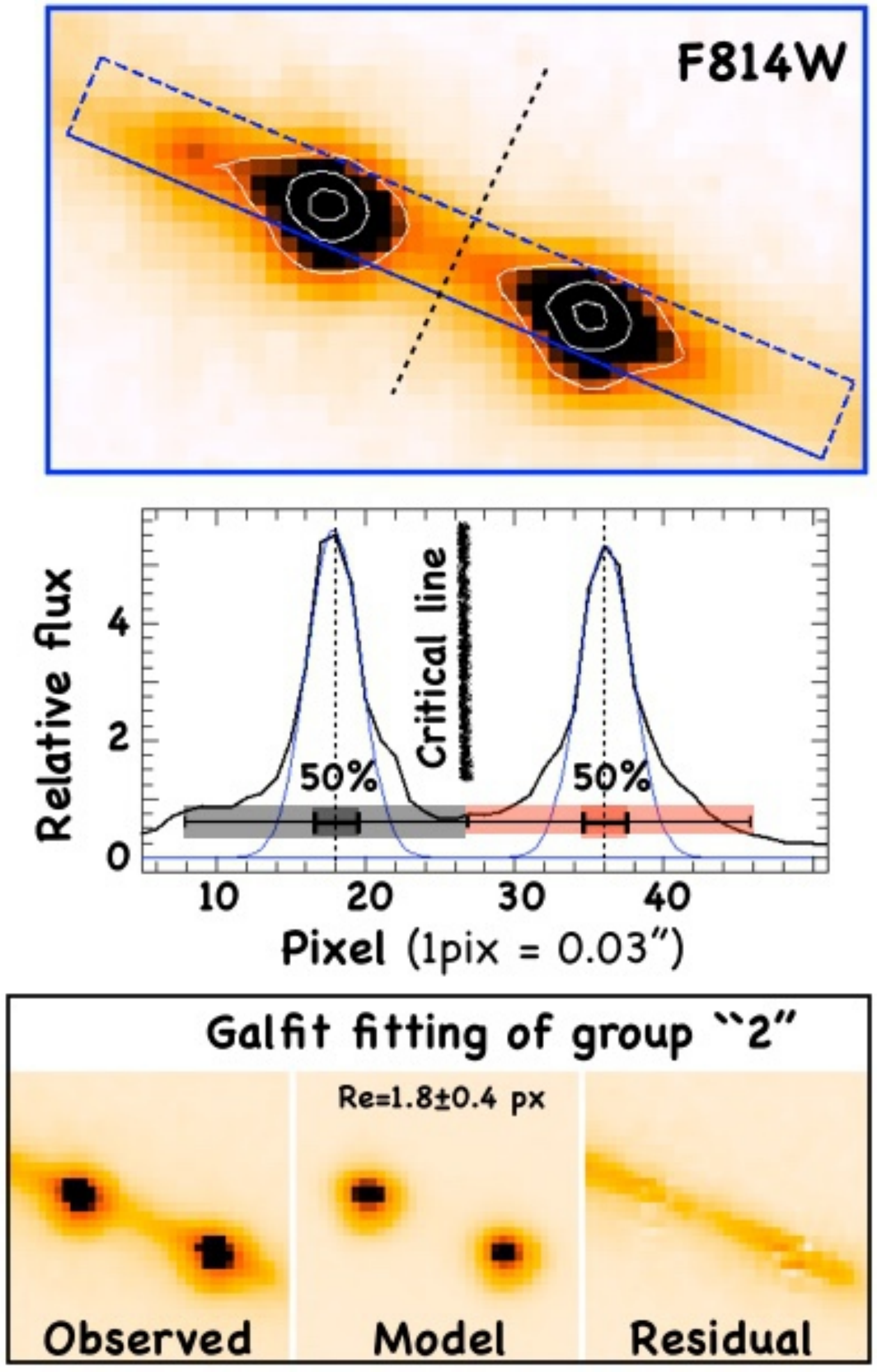} 
\caption{The top panel shows the F814W zoomed image of the multiple images ``2'' and ``3'' with the rectangular aperture 
($1.6'' \times 0.13''$) used to calculate the profiles reported in the middle panel, in which the 50\% area along the tangential direction 
is highlighted (shaded regions), and the FWHM ($0.13''$) of the F814W-band is superimposed with a blue line.
The expected crude position of the critical line is also indicated. The symmetry of the profiles (also outlined with the
white contours in the F814W images, in the top panel) despite the vicinity of the knots
to the critical line ($0.26''$) suggests the objects are quite compact. In the bottom 
the {\tt Galfit} solution in the same band is also shown, with Sersic index 4.0 and effective radius $\lesssim 2$ pixel (see text for details).}
\label{profiles}
\end{figure}

Figure~\ref{profiles} shows the two multiple images ``2'' and ``3'', each one located $0.26''$ away from the critical line, 
that necessarily falls between them.
The giant arc$-$like shape also implies the magnification is mainly tangential,
such that the total magnification is close to the tangential one, $\mu_{TOT} \simeq \mu_{tan}$
(see discussion in \citealt{vanz17b}).
Despite the knots lie in a region with a potentially
steep tangential magnification gradient, 
the light profiles along the same direction are symmetric (see Figure~\ref{profiles}, middle panel),
strongly supporting the fact that the object is intrinsically compact
(see also discussion by \citealt{vanz16a} on another similar case). 
As shown in the middle panel of the same figure, the LyC$-$knot is marginally resolved in the
HST/F814W image, close to the resolution in that band, in which the PSF FWHM is $\simeq 0.13''$ (though it is not a point-like source).
Indeed, specific {\tt Galfit}-based deconvolution analysis (as similarly performed in \citealt{vanz16a,vanz17b}) 
produces effective radii ($R_e$) of the order of $1.0-2.0$ pixel, for a Sersic index in the range {\it n} = $0.5 - 5$. 
It is worth noting, however, that a large {\it n} and $R_e$ would produce non symmetric tangential profiles, as mentioned above.
This will be fully investigated with dedicated simulations of all the knots and the emerging light profiles 
by placing objects with known structural parameters in the source plane close to the caustics (see appendix A of \citealt{vanz17b}), 
once the lens model will be developed.

\section{Measuring the significance of the emission lines detected in the X-Shooter spectrum}
\label{scan}

The statistical significance of the emission lines reported in Table~\ref{tab} has been calculated performing spectral scans over the
reduced spectra in the UVB, VIS and NIR arms. In particular, we describe here the case of the VIS arm, in which we explore the presence
and broadness of the  \heii\ line.

\begin{figure*}
\centering
\includegraphics[width=17.5cm]{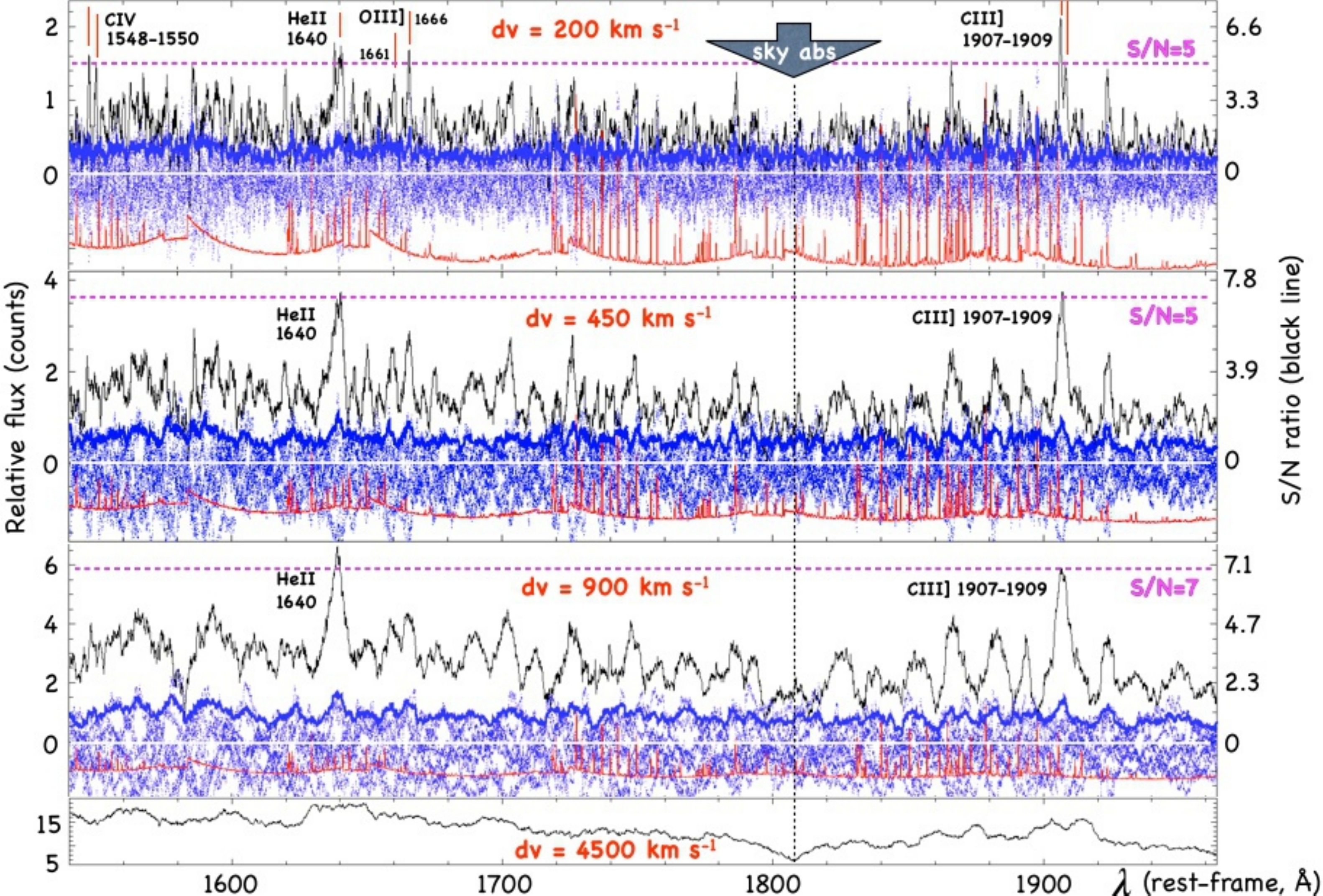} 
\caption{The spectral scan performed in the VIS arm with the aim to quantify the statistical significance of the spectral features of {\em Ion2}  reported in
Table~\ref{tab}. As discussed in the text the scan is performed by using four windows, all of them having the same spatial scale (of $0.8''$) and
different velocity widths {\em dv} = 200, 450, 900 and 4500 \kms. The black lines show the scan calculated over the target trace. 
The small/large spectral features are detected with small/large {\em dv}, e.g., the narrow doublets and the \heii\ emission, respectively. 
The same scan performed in the region of the spectrum avoiding the position of the target provides an estimate of the 
error fluctuation in the chosen window (cloud of small blue dots), from which the standard deviation is also extracted and indicated with a thick, nearly flat, 
 blue line. Therefore, the black and the thick blue lines represent the signal and the error, respectively. The Y-axis on the right 
 indicates the S/N ratio, with the values at S/N = 5 and 7 marked with magenta dashed lines. The position of the sky emission lines and the shape due to
the various orders within the VIS arm are shown in red, suitably rescaled for the figure purposes. The 7600\AA~sky absorption band (corresponding to
rest-frame 1800-1820\AA) is also indicated with a large arrow and is consistent with the dip visible in the continuum of the target in that position.
In the bottom panel, {\em dv} = 4500 \kms, such continuum dip is clearly recovered.}
\label{SN}
\end{figure*}

A spectral scan over the two-dimensional reduced spectrum is performed by using a widow with spatial scale {\em ds} and velocity {\em dv}.
{\em ds} has been fixed to $0.8''$, slightly larger than the mean seeing during the observations (the spatial scale in the VIS arm is $0.16''$/pix).
Thee velocity widths have been used, {\em dv} = 200, 450 and 900 \kms. Small(large) {\em dv}  captures small(large) spectral features.
The scan has been performed pixel by pixel in the wavelength direction (0.2\AA/pix), while the spatial direction windows do not
overlap to each other (they are independent). Moreover, we distinguished between the position of the target (that lies at a fixed position) 
and the rest of the slit. After excluding the target and the edges of the slit we end up with 10 independent windows corresponding to
each wavelength position. Figure~\ref{SN} shows the results of this exercise. The black line in each panel shows the scan performed at the
position of the target. The scan with small {\em dv} recovers the narrow features and doublets discussed in the text, 
like the \civ, \oiiiuv, \heii\ and the \ciiidoub\ lines. The results of the same scan avoiding the target is shown with blue dots 
which are more distributed around the zero value and follow the spectral pattern of the sky emission (shown with red line and rescaled as a guidance).
The thick blue line represents the standard deviation of the cloud of blue points calculated over the 10 windows available at each wavelength position.
The black and the thick blue line are the signal and the error, respectively. The S/N ratios reported in Table~\ref{tab} are inferred from this analysis.

The same spectral scan has been computed by enlarging {\em dv}. Figure~\ref{SN} clearly shows that as {\em dv} increases the tiny 
spectral features disappear, hidden by the continuum fluctuation. The continuum shows a small (but wide) dip at $\lambda \sim 1810$\AA, that
corresponds to 7620\AA\ at the observed frame, due to the well known sky absorption band (7600-7640\AA). 
Two main spectral features associated to {\em Ion2} stand out from the continuum:  the \ciiidoub\ doublet that is detected as a single 
entity and the broad emission identified at the position of \heii, with a S/N $\simeq 4$.

\end{document}